# From single phase to dual-phase TRIP-TWIP titanium alloys for the improvement of the yield strength

L. Lilensten[1,2], Y. Danard[2], R. Poulain[2], R. Guillou[3], J. M. Joubert[4], L. Perrière[4], P. Vermaut[2,5], D. Thiaudière[6], F. Prima[2]

[1] Max-Planck-Institut für Eisenforschung GmbH, 40237 Düsseldorf, Germany
[2] PSL Research University, Chimie ParisTech, Institut de Recherche de Chimie Paris, CNRS UMR 8247, Paris, France
[3] DEN-Service de Recherches Métallurgiques Appliquées (SRMA), CEA, Université Paris-Saclay, F-91191, Gif-sur-Yvette, France
[4] Université Paris-Est, ICMPE (UMR 7182), CNRS, UPEC, 2-8 Rue Henri Dunant, 94320, Thiais, France
[5] Sorbonne Universités, UPMC Université Paris 06, UFR926, 75005 Paris, France
[6] Synchrotron SOLEIL, L'orme des merisiers, Saint-Aubin, F-91192 Gif-sur-Yvette, France

**Abstract**

Aiming at increasing the yield strength of transformation and twinning induced plasticity (TRIP and TWIP) titanium alloys, a dual-phase α/β alloy is designed and studied. The composition Ti – 7 Cr – 1.5 Sn (wt.%) is proposed, based on an approach coupling Calphad calculations and classical Bo-Md design tool used in Ti-alloys. Its microstructure is made of 20% of α precipitates in a β matrix, the matrix having optimal Bo and Md parameters for deformation twinning and martensitic transformation. The alloy indeed displays a yield strength of 760 MPa, about 200 MPa above that of a Ti – 8.5Cr – 1.5Sn (wt.%) single β phase TRIP/TWIP alloy, combined with good ductility and work-hardening. In situ synchrotron X ray diffraction and post-mortem electron back-scattered analyses are performed to characterize the deformation mechanisms. They evidence that the TRIP and TWIP mechanisms are successfully obtained in the material, validating the design strategy. The interaction of the precipitates with the {332}<113> β twins is analyzed, evidencing that the precipitates are sheared when hit by a twin, and therefore do not hinder the propagation of the twins. The detailed nature of the interaction is discussed, as well as the impact of the precipitates on the mechanical properties.

1. Introduction

In recent years, a threshold has been overcome in β-titanium alloys with the quick emergence of a new generation of alloys displaying combined deformation mechanisms such as transformation induced plasticity (TRIP) and twinning induced plasticity (TWIP) effects. Thanks to these two effects; unprecedented ductility and superior work-hardening are obtained together for these alloys, opening the way for a broader field of applications [1–4]. As dislocation glide controls the deformation in stable β (bcc) Ti-alloys, the two alternative deformation mechanisms may be obtained by destabilizing the bcc phase in a controlled way [5]. The addition of α-stabilizing alloying elements, or the decrease in the β-stabilizers concentration allows to bring the alloy to a metastable region, where its Ms temperature is located just below room temperature [6–10]. Therefore, upon quenching from the β-region, a 100% metastable β structure can be retained. Depending on its level of metastability, the later will display the TWIP mechanism, or the TRIP one, or a combination of both. Several strategies have been proposed to predict the deformation mechanisms obtained for a nominal composition. They go from very specific DFT calculations to describe the systems [11], to more "ready to use" techniques [12], such as the Bo-Md $d$-electron alloy design which will be described in more details later on [1,13]. Thanks to all the available design techniques, an increasing number of TRIP/TWIP β-metastable alloys have been proposed and studied recently [1,2,14–16]. The understanding of the relationship between the mechanical properties and the deformation mechanisms is thus getting better, even though the interplay between twinning, phase transformation and dislocation glide is highly complex, and makes the hypothesis of a single framework applicable to every TRIP/TWIP alloy unlikely [14,17].

Balancing the numerous qualities of this family of alloys, their ductility and work hardening, but also their low density for example, the yield strength remains a potential issue for some applications, such as the aerospace related ones. Indeed, rather low yield strength values are generally obtained in most of the developed systems, often below 600 MPa [1–3,14]. Some strategies have been suggested, aiming at increasing the yield strength. Following the Hall-Petch principle [18], reducing the grain size decreases the dislocation mean free path, and therefore increases the yield strength. However, studies have shown that if grain size reduction does increase the YS in TWIP alloys, it also makes twinning more difficult, resulting in a subsequent decrease of the work hardening rate and the ductility [19]. Similar results have been obtained for TRIP alloys, and this deformation mechanism may even be suppressed [20]. Another possible route is to use the solid solution effect to harden the material. This has been investigated in a few studies, but, up to now, without tremendous increase of the yield strength, or with a strong decrease of the work-hardening rate and/or the ductility [6,21–24]. Finally, some studies focused on increasing the yield strength by using precipitates to reinforce the matrix. Controlled ω nano-precipitation has proven to be rather efficient and this strategy has been receiving increasing attention recently, though the relationship between precipitates, work hardening and ductility is not straightforward [25,26]. Following this precipitation strengthening strategy, we propose to design a dual-phase TRIP-TWIP alloy, with a β-metastable matrix reinforced by α precipitates. Deformation by these two mechanisms have been observed in dual phase systems [27], and a recent study indeed showed that precipitation of 4% of α could help increase the yield strength without loss of ductility or work-hardening, thanks to the TRIP and TWIP mechanisms [28]. Their great mechanical properties are however explained by the compositional change of the matrix, the small grain size and the large number of low-angle grain boundaries; the influence of α is rather confined to the prevention of the β grain growth, as well as the formation of a dense network of low angle grain boundaries. Other dual-phase β-metastable Ti-alloys have been claimed before, but to the authors best knowledge, they mostly concern β-metastable alloys in which α is precipitated, and where the TRIP and TWIP properties of the β-matrix are lost [27,29]. In this work, the two phases are co-designed to bring specific properties to the alloy, *i.e.* the ductility and work-hardening for the β-matrix, and the yield strength from the α-precipitates. This proof of concept aims at opening the way to a new generation of tunable TRIP/TWIP Ti alloys, beyond single phase materials. The design strategy, combining thermodynamical calculations using the Calphad method, and the Bo-Md alloy design tool, is first described. The mechanical properties and deformation mechanisms of the alloy are then studied.

## 2. Materials and methods

Calphad calculations were performed using the Thermo-Calc software and the Cr-Sn-Ti database published by Gao *et al* [30]. Buttons of 200g were prepared by arc melting, and subsequently forged and hot rolled, to obtain square bars. They were solution treated at 1173 K for 900 s (in the β field) under air, followed by water quenching. Specimen were then cold rolled to a final thickness of about 0.65 mm, corresponding to a thickness reduction around 80%. The cold-rolled specimens were heat treated under high purity Ar at 1103 K for 1800 s (heating rate of 20 K/min) and water quenched to reach a fully recrystallized β structure. The α-phase was then precipitated by a thermal treatment at 1018 K for 1800 s (heating rate of 20 K/min) under high purity Ar, followed by water quenching.

Tensile specimen with dog bone shape and gauge dimensions of 50 x 4 x 0.65 mm$^3$ were prepared by wire Electro Discharge Machining. Tensile tests were performed at room temperature, at a strain rate of 1.7 x 10$^{-3}$ s$^{-1}$ on an Instron machine equipped with a 10 kN load cell and with an extensometer with a gauge length of 10mm. Samples for microscopy were grinded on SiC papers and finally polished with an OPS-H$_2$O$_2$ mixture.

Back scatter electron (BSE) imaging, to characterize the initial microstructure, was done using a Zeiss Leo1530 field emission gun scanning electron microscope (SEM) operated at 20 kV. Images were treated for quantification of the α-phase using the Image J software.

Specimens for electron back-scattered diffraction (EBSD) analyses were electropolished at 278 K with a solution of 2-butoxyethanol, methanol, perchloric acid and hydrochloric acid. EBSD analyses were performed on a Zeiss-Crossbeam XB 1540 Focus Ion Beam FIB-SEM instrument equipped with a field emission gun operated at 15 kV. The step size was comprised between 30 and 50 nm. EBSD data collection was done by an EDAX/TSL acquisition system, and the results were then analyzed using the TSL/OIM analysis package. Energy dispersive X ray diffraction (EDX) line profiles were measured across precipitates using a Zeiss Merlin FEG-SEM operated at 20kV.

*In situ* SXRD (synchrotron X ray diffraction) analyses were carried out at the DiffAbs (diffraction and absorption) beamline at Soleil, the French synchrotron facility. XRD measurements were performed in reflection mode with an energy of the X-ray beam fixed to 12.475 keV ($\lambda=0.994$ Å). Using this energy combined with an incident angle of 9°, the penetration depth was approximately 20 µm. A 600 x 300 µm² x-ray beam size (HxV fwhm) at the center of the 6-circle diffractometer was used in order to obtain good statistics of grains during measurements. The diffraction patterns were recorded thanks to a bi-dimensional XPAD-S140 detector (hybrid pixel technology, 560 x 240 pixels² image size) [31,32]. The detector was mounted on the 2θ goniometer arm with a detector sample distance of 580 mm (granting an angular resolution of 0.013°). After recording, each diffraction pattern was converted to diffraction diagram (radial integration) with a typical [32°-50°] 2θ angular range. A Proxima-Micromecha tensile machine with a 3kN load cell was used for *in situ* tensile tests with specimen's dimensions of 50 x 2.95 x 0.49 mm and a gauge length of 20 mm.

3. Results

    a. Alloy design approach

The approach proposed here is to reinforce a TRIP/TWIP matrix with hardening precipitates. Care must be taken to consider the precipitation of the second phase in the determination of the alloy's nominal composition, and therefore anticipate the compositional change of the matrix. For this study, the precipitation of α particles inside the β-grains is proposed and will induce chemical partition of the elements. The determination of the composition was done based on the following reasoning: the microstructure prior deformation should be made of about 20% of α-precipitates, in a β-matrix, in order to keep β as the main phase that controls the deformation mechanisms, and have enough α precipitates to harden the microstructure. Thus, after precipitation of 20% of α, the β matrix should have an optimum composition with adequate Bo-Md parameters for TRIP/TWIP.

Establishing the alloy's nominal composition is done by coupling two techniques: Calphad is used to calculate the composition of the α and the β phases for a given nominal composition and microstructure, and the Bo-Md technique to assess the TRIP/TWIP character of the β phase. The Bo-Md tool was first developed by Morinaga *et al.* for superelastic Ti alloy design [13,33]. In this method, the average bond order (Bo) and average energy of the *d*-orbital (Md) of an alloy's composition are calculated. They reflect the cohesion strength and the chemical stability of the alloy, respectively. These theoretical data, extrapolated from molecular orbital calculations, are coupled with a stability map based on experimental results [13]. That allows to draw the limits of stability for the β-phase, and then to target the instability region for the TRIP/TWIP alloy design. This method has been proven to be efficient and easy to use [1,2,14–16].

According to recent results, the single phase Ti – 8.5Cr – 1.5Sn (wt%) alloy, designed by the Bo-Md method, displays a remarkable work-hardening (with an ultimate tensile strength above 1200 MPa), a uniform deformation of 0.36, and a yield strength of 550 MPa [4], based on intense TRIP/TWIP combined effects. Furthermore, the thermodynamical properties of the Ti-Cr-Sn system have been published and could be used in this study for the Calphad calculations [30]. The Ti-Cr-Sn system is therefore an excellent candidate for this study, and Ti–8.5Cr–1.5Sn (wt%) will be considered as the "reference alloy" in the following, in order to design a "dual-phase" grade based on this system.

The design results are illustrated in Figure 1. Considering a nominal composition Ti-7Cr-1.5Sn (wt.%) (point 1 in Figure 1a), Calphad predicts that, for a microstructure containing 20% α (points 2 in Figure 1a), the β-phase has the composition Ti-8.6Cr-1.4Sn (wt.%), very close to the targeted composition of the reference alloy. As for the α precipitates, they would have the composition Ti-0.5Cr-2.1Sn (wt.%): with no surprise, Cr, which is a β stabilizer, is rejected from the precipitates to the matrix. One can note that the C15 phase is predicted by the calculation, however it has not been observed experimentally. Constrained calculations omitting C15 do not show any difference for the bcc and the hcp phases compositions and temperatures, in the α+β domain. The stability and expected deformation mechanisms following the Bo-Md design are illustrated in Figure 1b. The alloy with nominal composition Ti-7Cr-1.5Sn (wt.%) (point 1 in Figure 1b) sits in the blue area, that deforms preferably by formation of stress induced martensite (TRIP) if the alloy is loaded in its fully β microstructural state. The Bo and Md parameters of the β-phase after precipitation of α are indicated by a light blue dot (see also point 2 in Figure 1b). One can see that it is very close to the targeted Ti-8.5Cr-1.5Sn (wt.%) reference alloy, in dark blue in Figure 1b.

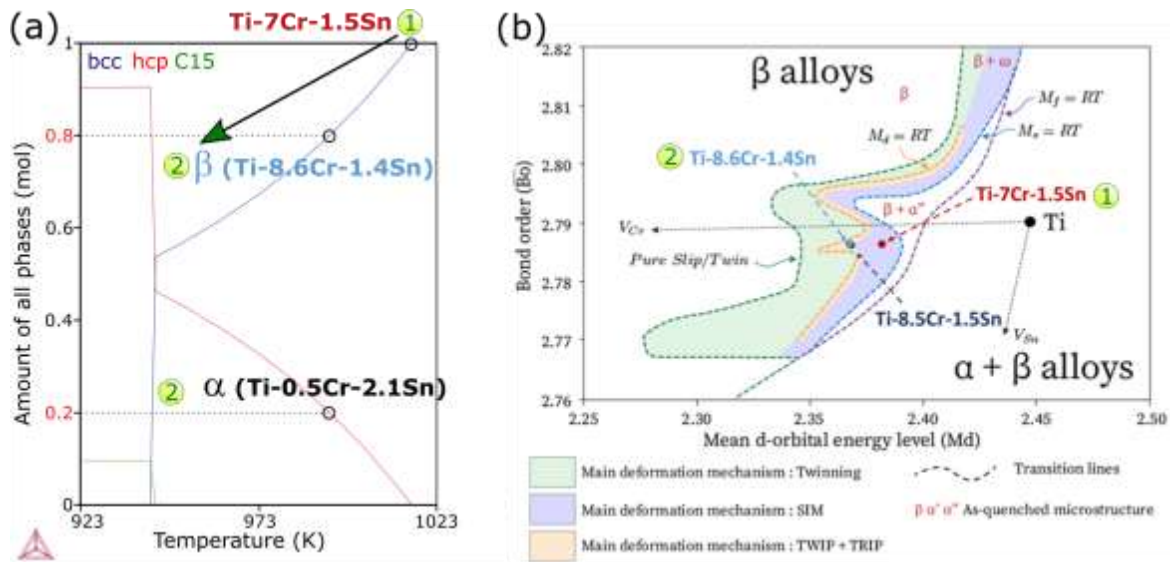

*Figure 1: (a) Calphad predictions for the designed nominal composition (1) Ti-7Cr-1.5Sn (wt.%): amount of the predicted phases (bcc-β, hcp-α and C15) as a function of the temperature. The composition of the bcc and hcp phases for a microstructure (2) containing 80% of β and 20% of α are given. (b) Bo-Md predictions for the nominal*

*composition (1) Ti-7Cr-1.5Sn (wt.%) and its β-phase (2) after precipitation of 20% of α, compared to the reference alloy Ti-8.5Cr-1.5Sn (wt%)* [4].

Based on these design results, the dual phase Ti-7Cr-1.5Sn (wt.%) alloy involving a α volume fraction of 20% is expected to have the desired properties. The alloy is thus prepared and studied in the following.

### b. Initial microstructure

The alloy with nominal composition Ti-7Cr-1.5Sn (wt.%) is prepared, based on the design results. After thermomechanical treatments, a recrystallized β microstructure is obtained by a heat treatment in the high temperature domain (point (1) in Figure 1a), and the α phase is precipitated in the β grains by a second thermal treatment in the α domain (points (2) in Figure 1a). The adequate experimental temperatures to obtain the desired microstructure were slightly off compared to the Calphad predictions. The obtained microstructure is represented in Figure 2. On this scanning electron microscopy (SEM) image taken with BSE contrast, the α-precipitates appear dark, which indicates that they are depleted in heavy elements. As Sn is expected to partition only slightly between the two phases, and Cr is expected to be massively rejected towards the β-matrix, the observed contrast is consistent with the expected Cr trend. Qualitative EDX analyses given in supplementary material confirm the depletion of Cr and enrichment of Ti at the α precipitates.

One can see that α is precipitating mostly as platelets inside the grains, with a homogeneous repartition. This phase is also found at the grain boundaries of the β grains, which is a classical observation in β-Ti alloys, where it forms $α_{GB}$ (elongated precipitate between the two grains) and $α_{wGB}$, characterized by precipitates aligned rather perpendicular to the grain boundary.

An average amount of 20% of α-phase is quantified by image analyses, by averaging the amount calculated for several images. This value is in perfect agreement with the design strategy.

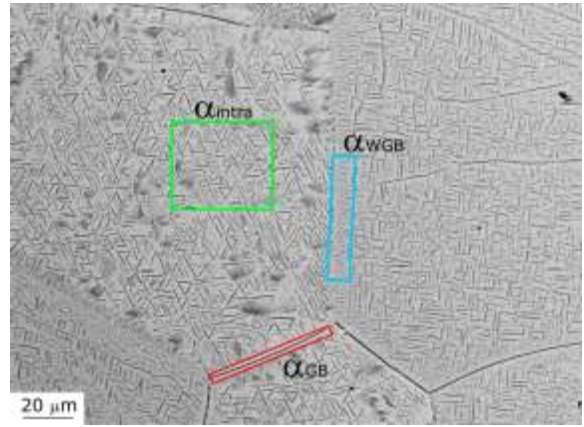

*Figure 2: BSE image of the microstructure of the alloy prior deformation. The α precipitates appear in dark in the β matrix.*

EBSD analyses show that the α-precipitates are found to obey the Burgers orientation relationship (BOR) which corresponds to $\{110\}\beta//\{0001\}\alpha$ and $<1\text{-}11>\beta//<11\text{-}20>\alpha$. Several variants are observed in each grain. As for the $\alpha_{GB}$, they usually fulfil the BOR with one of the neighboring grains.

### c. Mechanical properties

The true stress – true strain tensile curve and the work-hardening rate curve of the new alloy, given in Figure 3, are compared with the reference alloy Ti-8.5Cr-1.5Sn (wt.%) (100% β TRIP/TWIP alloy) [4].

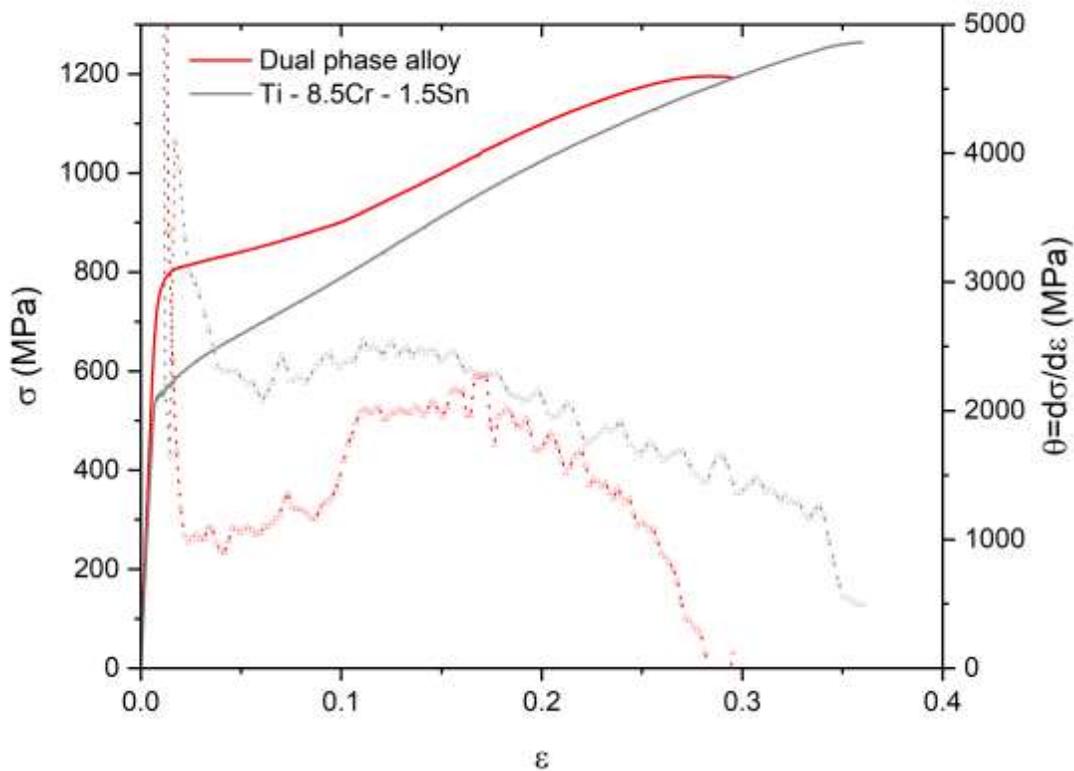

*Figure 3: True stress σ – true strain ε tensile curves (solid lines) and work-hardening rate θ (dashed lines) of the newly designed dual-phase alloy (in red) and the reference 100% β alloy (in grey)* [4].

The dual phase alloy displays an elastic modulus of 95 GPa, a very high yield strength of 760 MPa, as well as a uniform elongation strain of 0.29 and an ultimate tensile strength of 1200 MPa. In spite of a small loss of ductility and work-hardening, compared to the reference alloy, an increase of the yield strength larger than 200 MPa is obtained. This result itself validates the strategy consisting in precipitating a secondary phase to reinforce the matrix. The dual-phase alloy displays a marked increase of the work-hardening rate around 0.15 strain, following a noticeable stress-plateau, that is commonly observed in TRIP alloys [34], suggesting the occurrence of this deformation mechanism in the new alloy. Comparing the intensity of the hump in the work-hardening rate curves, it can be suggested that the TRIP effect is stronger in the dual-phase alloy (bigger hump) than in the reference alloy.

### d. Deformation mechanisms

Strain induced martensite formation was characterized by *in-situ* synchrotron X ray diffraction. The intensity in the X ray diffractograms corresponds to the integration of the intensities measured on the whole length of the XPAD-S140. Figure 4 shows the evolution of the diffraction patterns as the strain increases, from the initial state up to 16%. At 16%, the sample was unloaded and a diffractogram was recorded in the unloaded condition. Diffractograms recorded at larger deformation levels are not displayed here, as they do not bring additional information, and furthermore become less defined due to the very large strains accommodated in the microstructure. The diffractogram at 0% strain shows the presence of the α phase and the β phase only, with lattice parameters $a=2.933$ Å and $c=4.669$ Å for α and $a=3.231$ Å for β. Due to the very large grain size of the β grains (>100 μm) in the initial state compared to the beam size, artifacts are observed on the 0% strain diffractogram, taking the shape of double peaks for the plan (002) at ~36°. This effect disappears as soon as the deformation starts, with subdivision of the grains into smaller domains linked to the deformation features (see supplementary materials for more details). At 4% strain, peaks that can be attributed to the stress induced martensite are already seen, such as the ones corresponding to the (112) and the (022) planes. This confirms that the TRIP mechanism is effective in this alloy, as was suggested already by the shape of the tensile and work-hardening rate curves. As the deformation increases, the intensity of the α'' phase increases as well, such as the (130) one. However, the larger strains accumulated in the structure broaden the peaks that become less resolved. When unloading, some of the martensite seems to reverse back, as would indicate the decrease in intensity of its (130) peak, and a new peak corresponding to the $(200)_{\alpha''}$ reflection appears.

One can also note the appearance of a peak around 41.5°. This does not match any of the possible reflections for the considered α, α'' and β phases, but it is consistent with the $(002)_\omega$ reflection of the ω phase, considering that ω originates from β, and fulfills the following relationships: $a(\omega) = \sqrt{2}a(\beta)$ and $c(\omega) = \frac{\sqrt{3}}{2}a(\beta)$ [35,36].

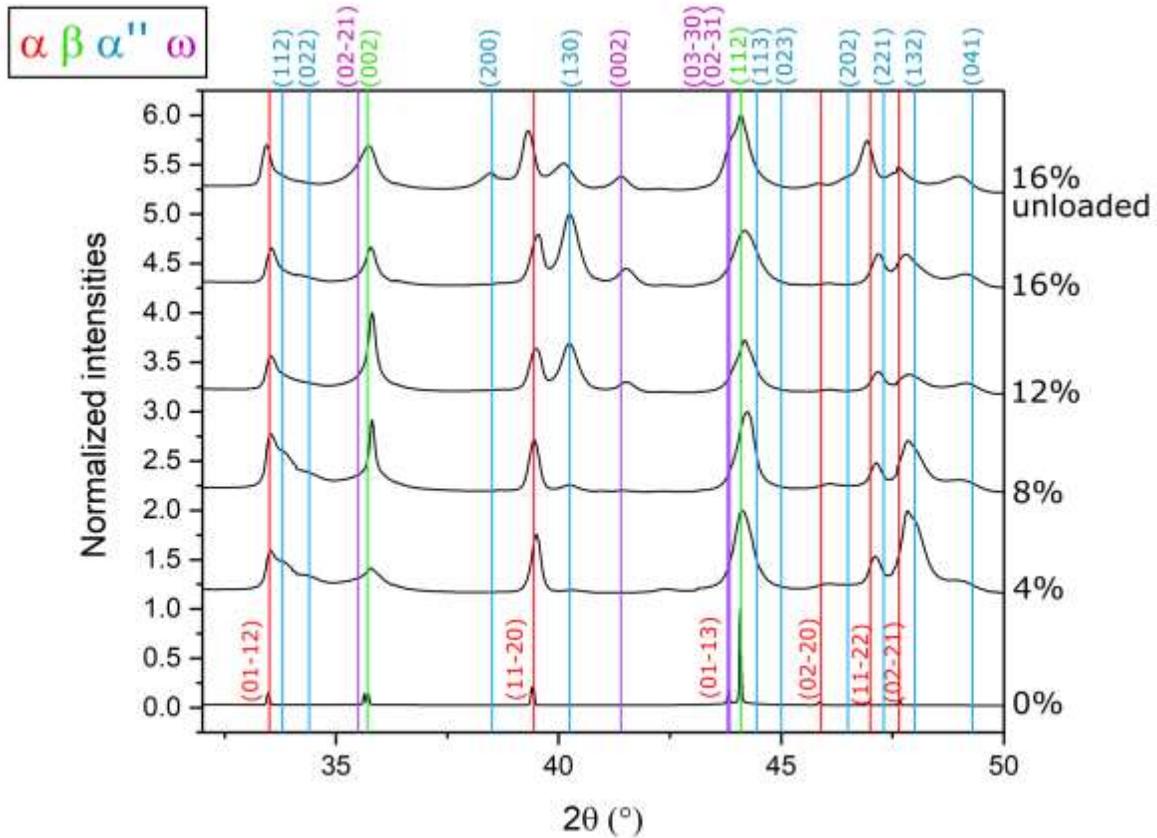

*Figure 4: synchrotron XRD patterns of the alloy between 0% and 16% strain, and after unloading at 16%. The diffraction reflection positions are labeled in red for the α phase, in green for the β phase, in blue for the α'' martensite and in purple for ω. Note that the vertical bar indicating the position of the (03-30) ω reflection hides that of the (01-13) reflection of α.*

Investigations of the deformation mechanisms was further completed using the EBSD technique ("post-mortem" observations). Due to the proximity in the diffraction patterns of the α hcp phase and the α'' martensite (orthorhombic), the indexing of the patterns is done only with the α-phase and the β-phase. Therefore, identification of the martensite with this technique is not possible.

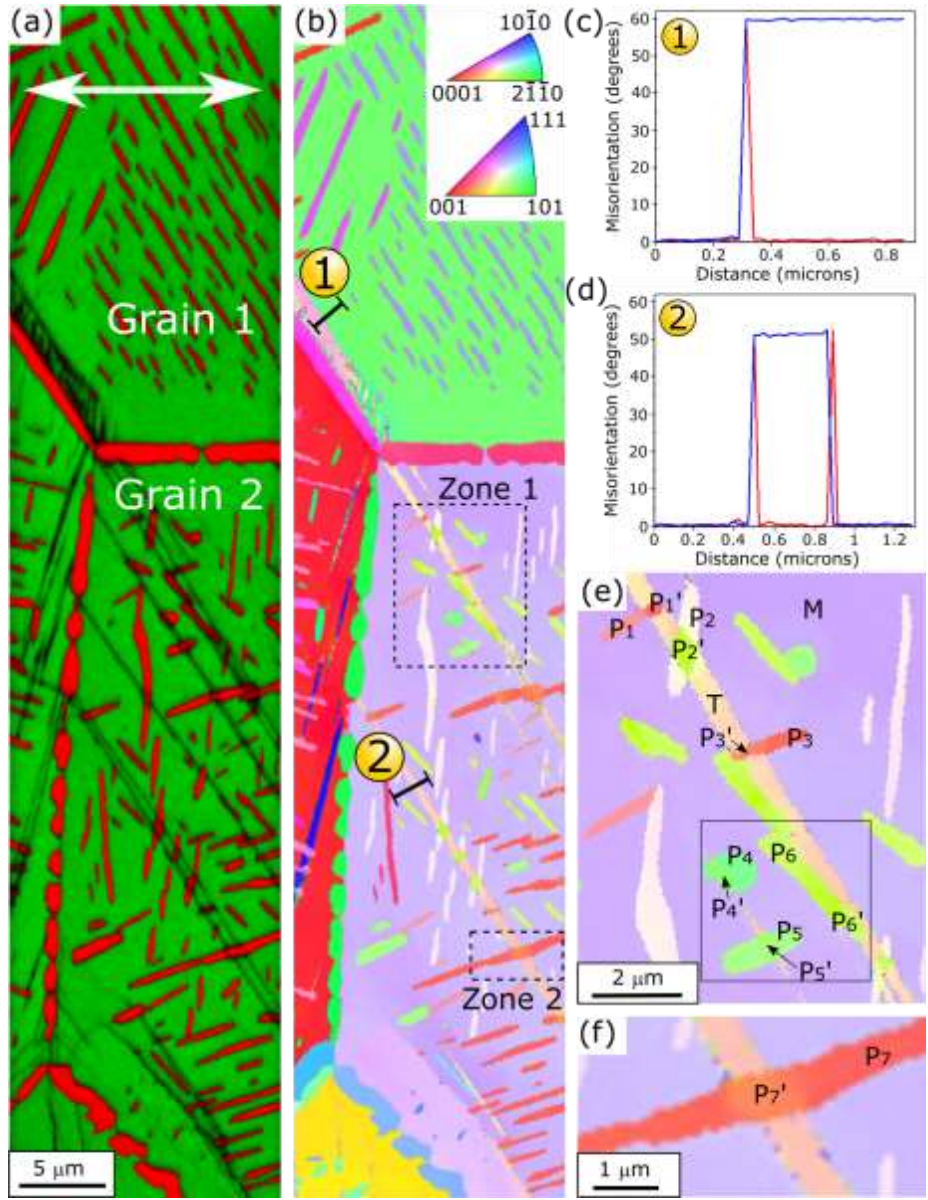

*Figure 5: EBSD analyses of a sample after 5% strain and unloading. (a) phase map over imposed with the quality index. α is in red and β in green. The tensile direction is indicated by the white double arrow. (b) Orientation map in the direction normal to the sample surface. (c) Orientation profile along the black line labelled 1 in (b). The point to point misorientation is plotted in red and the point to origin misorientation is plotted in blue. (d)*

*Orientation profile along the black line labelled 2 in (b). The point to point misorientation is plotted in red and the point to origin misorientation is plotted in blue. (e) Zoom into the zone 1 inset of (b). (f) Zoom into the zone 2 inset of (b).*

EBSD analyses of a 5% deformed microstructure after unloading are displayed in Figure 5 (tensile direction indicated by the white double arrow in Figure 5a). The phase map in Figure 5a confirms a surface fraction of α phase of 20%. The index quality (IQ), over imposed on the phase map, reveals the presence of deformation bands in the grains (lower IQ). The orientation map is given in Figure 5b. The misorientation profiles (indicated by (1) and (2) in Figure 5b), across two deformation bands (see arrows) are measured and plotted in Figures 5c and 5d respectively. The profile 1 (Figure 5c) highlights a misorientation of 60° corresponding to a β {112}<111> twin. This twin is observed at the interface between $α_{GB}$ and the β grain. Orientation analyses indicate that the BOR is not fulfilled between $α_{GB}$ and the matrix in this case, nor it is between the twin and $α_{GB}$. The IQ shows that this area is not so well indexed, which can be explained by the local strain. The second deformation profile (Figure 5d) highlights a misorientation of 51°, and this deformation band was identified as a {332}<113> β twin. Several {332}<113> twins with the same orientation are observed in grain 2. Other deformation bands identified as {332}<113> twins are found, such as in the red neighboring grain (blue orientation in Figure 5b). Therefore, it is confirmed that the β-matrix of this dual-phase alloy clearly displays TWIP properties additionally to the TRIP effect observed on SXRD patterns.

Figures 5e and 5f are magnified views of the insets labeled Zone 1 and Zone 2 in Figure 5b. They display the interaction of a twin with α-precipitates of 7 different orientations ($P_1$ to $P_7$) corresponding to different variants. All the precipitates are sheared by the twin: the precipitate $P_i$ transforms to a precipitate $P_i$' on the sheared zone. The general behavior observed here relates to a simple shearing of the precipitates by the mechanical twins, upon hitting them. It is worth noting that the twins trajectory remains quite straight even after successive interactions with α-precipitates, without visible deviation. The fact that the twins do not seem disturbed by the precipitates, and that this deformation mechanism is not hindered by the high density of precipitates, is remarkable.

## 4. Discussion

The alloy Ti – 7 Cr – 1.5 Sn (wt. %) was designed by coupling the electronic Bo-Md approach and the Calphad method, in order to obtain a dual-phase TRIP/TWIP alloy. Such alloy would benefit from the ductility and the work hardening of the β-matrix, when the precipitate could help increasing the stress, and especially the yield strength. It can be noted that the role of the precipitates is most likely more complex than a simple increase of the alloy's strength, especially considering the alloy's plasticity and the interplay of the various deformation mechanisms of the targeted alloy. After alloy preparation and processing, a microstructure made of 20% of α precipitates in the β matrix is obtained. The precipitates are found at the grain boundaries and inside the grains. They always fulfil the BOR with the matrix in the case of intragranular α, and the grain boundary α always fulfills the BOR with at least one of the two neighboring β grains. The mechanical properties obtained for the alloy reveal that from a mechanical point of view, the design strategy was successful, since a very good elongation and a good work-hardening are obtained, along with an increase in yield strength larger than 200 MPa. Study of the deformation mechanisms by *in situ* SXRD and post-mortem EBSD show that the deformation mechanisms of the dual phase alloy are in good agreement with the Bo-Md predictions. Orthorhombic martensite is forming from the early stages of the deformation. {332}<113> twins are observed in the grains, with several twinning variants activated at a strain of 0.05, and {112}<111> twinning was also observed, indicating that the two twinning modes contribute to the alloy's plasticity. Formation of ω phase at large strains observed with SXRD (Figure 4) could be correlated to the formation of the {112}<111> twins. Indeed, ω has already been observed to form at the interface between {112}<111> twins and the matrix [35,37–39]. Finally, upon unloading at 16% strain, a new reflection corresponding to (200)α'' appears on the XRD pattern. Taking into account the fact that, under load, this reflection was not detected, it can be suggested that some martensite also forms in the alloy during unloading, in order to accommodate internal strain incompatibilities.

The impact of α precipitates on the mechanical properties is complex. From the alloy design, the β matrix has the same composition in the reference and in the dual-phase alloy. Therefore, it is suggested that the differences regarding the plastic deformation (Figure 3) can be attributed to the α precipitates. They are discussed in the following, starting first with the interaction between the precipitates and the {332}<113> twins, and then, in a second time, with proposed explanations for the mechanical behavior.

### a. Interaction between the precipitates and the {332}<113> deformation twins
#### i. Background on the possible interaction and theoretical study

When a twin intersects a precipitate $P_i$, 3 theoretical situations may arise.

In the first case, there is a direct gliding system available for $P_i$ (aligned with the incident (332) β twinning plane, or close to it) and the precipitate deforms by dislocation glide. Dislocation storage in the precipitate may lead to strong strain fields, that can produce orientation gradients in between the sheared area and the rest of the precipitate. In the second case, the precipitate has a twinning system that matches the {332} twinning plane: on a stereographic projection, there should be a {332} plane close to a K1 hcp plane. That would imply that the {332} mirror plane of the matrix and the twin is parallel to a mirror plane for $P_i$ and $P_i$'. In the third case, the precipitate has no direct available system to transfer the shear carried by the twin into the precipitate, and reacts as an un-deformable object. This would lead to the formation of a strong strain field at the precipitate boundary corresponding to the backstress accommodated by the matrix.

The study of all the possible interactions between the β-twins and the α-precipitates can be reduced to that of one precipitate (fulfilling the BOR with the matrix) with the 12 possible {332}<113> twinning systems. Following the stereographic projections (see in supplementary materials), only one {332} bcc twinning plane matches perfectly with a hcp {11-20} plane, but such plane is not related to a dislocation glide system in hcp structures. No hcp twinning plane over imposes perfectly with a {332} bcc plane; in the best case, some, such as a {-1012} plane, are not so far away from a {332} plane. Therefore, the bcc twinning shear is supposed to be accommodated in the precipitate by combination of elasticity, to bring a hcp plane in a matching position, and subsequent glide or twinning.

## ii. Discussion on the experimental results

Deeper analysis of the experimental results on the interaction between twins and precipitates is performed. The area framed in Figure 5e is taken as an example, as it displays cases representative to what is observed for the other precipitates. More EBSD analyses are performed (11 twin/precipitates interactions studied, not displayed here) to confirm the occurrence of the effects described hereafter.

The orientation map of the considered area is given in Figure 6a. For a better color contrast, the orientation map is given along the tensile direction instead of the normal direction.

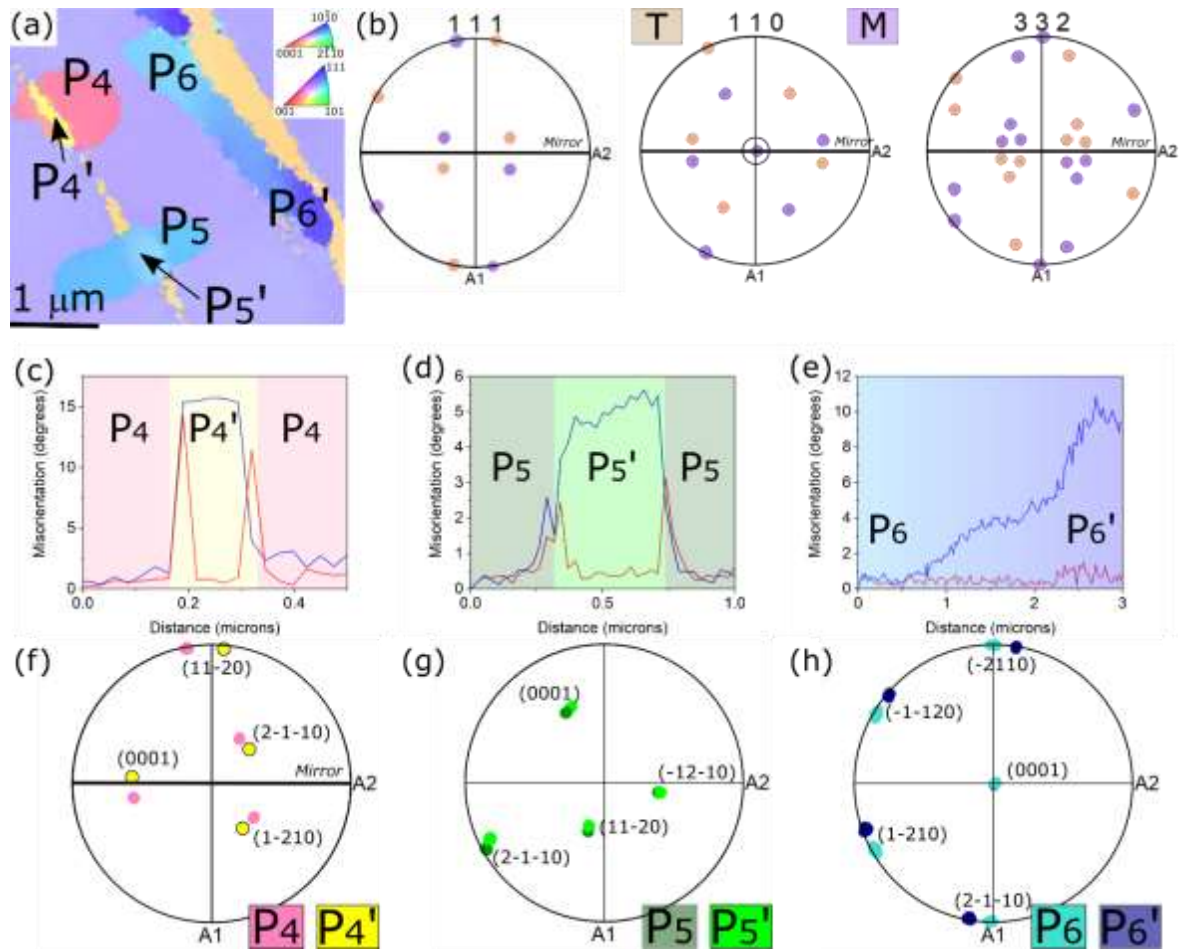

*Figure 6: (a) EBSD orientation map along the tensile direction of the framed zone of Figure 5e. (b) Stereographic projection of the matrix (lilac) and the twin (beige) for the 111, 110 and 332 poles, the sample being virtually rotated to have the mirror plane of the twin along the equator. The (110) plane common to the matrix and the twin is indicated by a black circle. (c), (d) and (e): Misorientation profiles across $P_4$-$P_4$', $P_5$-$P_5$' and $P_6$-$P_6$' respectively. The point to point misorientations are plotted in red and the point to origin misorientations are plotted in blue. (f), (g) and (h): Stereographic projections showing the (0001) pole and the {11-20} poles of $P_4$ and $P_4$', $P_5$ and $P_5$', and $P_6$ and $P_6$', respectively.*

Orientation changes are visible in the precipitates, confirming the shearing effect of the twin. As shown on the misorientation profiles plotted in Figures 6c to 6e, the misorientation angle (between $P_i$ and $P_i$') is rather small (5°) in the case of $P_5$-$P_5$', and larger for the other cases (10° and 15° for $P_6$-$P_6$' and $P_4$-$P_4$' respectively). It is also observed that the orientation change is very sharp for $P_4$ and $P_5$, but that there is an orientation gradient for $P_6$. This was always the case: when the habit plane of the precipitate is perpendicular to the twin plane, there is a sharp orientation change, whereas when the habit plane of the precipitate is rather parallel (or close to) the twin plane, the orientation changes gradually. It is proposed that the abrupt interfaces are found when the incident shear is very localized, confined to a narrow area at the twin/precipitate interface.

To understand further the shearing behavior of the precipitates, stereographic projections for the twin/matrix and the precipitates are given in Figures 6b, 6f, 6g and 6h; with the sample virtually rotated to bring the 110 pole common to the twin and the matrix to the center of the stereographic projection, and to bring the (332) twinning plane at the equator of the projection, *i.e.* with its pole at the north and south poles.
They first allow to confirm that all the precipitates $P_4$, $P_5$ and $P_6$ are in BOR with the matrix, but regarding their sheared variants, only $P_6$' fulfills the BOR with the twin, when $P_4$' is close to the BOR with the twin and $P_5$' has no particular orientation relationship with the twin (for more details see Figure S4 in the supplementary materials). But they also allow to see a new relationship between $P_4$ and $P_4$': the mirror plane that transforms the matrix into the twin also applies to P$_4$, that is transformed into P$_4$' (Figure 6f). However, regarding P$_5$ and P$_6$, no such relationship is found (Figure 6g and 6h). It can be noted that P$_6$ and P$_6$' are closer to a mirror relationship, that could be reached by a small rotation around (0001), when it would require a very large rotation of P$_5$' to be put in a mirror relationship with P$_5$.

A mirror relationship between P$_i$ and P$_i$' is observed in 3 cases out of 11 studied ones (see table 1), for various orientations of the precipitate with respect to the twin - with its (0001) plane parallel to the (110) plane common to the twin and the matrix or not. However, if this mirror plane characterizes a twinning relationship, it does not correspond to the classical hcp twins. The twinning plane could not be identified within the possible low Miller indices planes. Analysis of the orientation of the other planes relevant for twinning or dislocation glide in hcp structures, *i.e.* {10-11}, {11-22}, {10-10}, {10-12}, {11-21}, {11-22} and {11-24}, does not show any match between these planes in P$_4$, P$_5$ and P$_6$, and the (332) bcc twinning plane.

Over 11 twin/precipitates interactions studied, all the precipitates are sheared when hit by a twin, and the additional orientation relationships detailed here happen to be representative of what happens which is summarized in table 1.

| P$_i$' fulfills BOR with the twin | Mirror plane | Misorientation between P$_i$ and P$_i$' | Occurrence from the 11 α precipitates studied |
|---|---|---|---|
| No | No | 5 to 10° | 5 |
| Yes | No | 10 to 13° | 3 |
| Yes | Yes | 13 to 18° | 3 |

*Table 1: Summary of the different possible shearing scenarii for a precipitate hit by a twin, depending on its orientation.*

Based on the knowledge on deformation of hcp structures and the present results, the way the precipitates exactly deform is, at the moment, unclear, and more work will be dedicated to unveil the detailed mechanisms. Yet, the fact that they are always sheared indicates a local rotation of the lattice, therefore the activation of several gliding systems, and also indicates that the precipitates are participating to the alloy's plasticity. Special orientation relationships are found after shearing of the precipitate (mirror relationship or BOR between $P_i$' and the twin) in more than half of the studied cases. Regarding the BOR relationship obtained between the twin and the sheared precipitate, it is obtained in 6 cases out of 11, suggesting that accommodation of the shear in the precipitate may be done with the additional objective of reaching BOR with the twin, which would decrease the interfacial energy between the twin and the sheared precipitate. The fact that the bcc mirror plane corresponds to a hcp mirror plane in 3 cases out of 11 may also suggest the activation of a high index twinning system that could not be identified. This may be favored by the very specific configuration observed here, with high amounts of shear carried by a twin that are localized only on a small part of the precipitate.

The measured misorientations, ranging from 5° to 18° are probably partly correlated to the shearing magnitude. However, P1, P2, P3 and P6 on the one hand and P4 and P5 on the other hand are sheared by the same twin (see Figure 5), but the misorientation measured in these precipitates is not the same. So the shearing provided by the twin does not only depend on the twin and the level of strain, but it is suggested that it may be normalized by the volume of the precipitate to shear, which would explain the different measured values. One can note that this would not apply for the "twinned precipitates" if the hypothesis of a high index twinning system was validated.

Shearing or engulfment of precipitates by/into deformation twins have already been observed in other structures, such as the transmission of twins between the γ matrix (fcc) and γ' precipitates ($L1_2$) in superalloys, or between the Mg matrix (hcp) and MnZn' precipitates [40,41]. However, in these cases, twin growth is carried by the glide of individual partial dislocations at the twin/matrix interface, each partial bringing the shear to twin the next atomic layer. If the transmission mechanism across precipitates is not well understood in Mg alloys, the coherent γ/γ' interface in superalloys allows an easier transmission of the dislocations from one side to the other. {332} twinning in bcc alloys is more complex, as the twinning mechanism itself is still debated [17,42,43]. Therefore, it is too early to better and fully understand the mechanism by which it shears the precipitate.

### b. Understanding the impact of the precipitates on the mechanical properties

The impact of the precipitates on the yield strength, the overall tensile curve, the alloy's strength and work-hardening is now discussed.

Regarding the increase of the yield strength, it may be considered that the latter is controlled by the critical shear stress required to nucleate twins and/or martensite, depending on which of the two mechanisms has the lowest critical shear stress. These two defects are considered to initiate at the grain boundaries, and then propagate in the grain's interior. In the present case, ribbons of α phase are precipitated at the former β/β grain boundaries. This may modify the local stress required to trigger the nucleation of a twin or of martensite, hence the increase in the yield strength. The grain size effect on the yield strength, through the Hall-Petch law, can be eliminated as the grain size of the reference alloy is 200 µm and the size of the recrystallized β grains of the dual phase alloy is approximately 120 µm, so both alloys have grains sizes of the same order of magnitude. Increase of the yield strength due to the precipitates following an Ashby-Orowan mechanism can also be discarded. Indeed, it can be estimated following the equation $\Delta\sigma_y = (0.538Gbf^{1/2}/X)ln(X/2b)$, where G is the shear modulus, *b* the Burgers vector, *f* the volume fraction and *X* the diameter of the particles [44]. G can be estimated from the elastic modulus *E* measured by the tensile test, following the *G/E=3/8* ratio commonly used for polycrystalline materials [45]. The Burgers vector is considered to be ***b***=1/2<111>, as classically in bcc structures, and equals to 2.80Å based on the lattice parameter extracted from the synchrotron XRD spectrum at 0% strain. The volume fraction is taken as 0.2, as expected from the alloy design, and the particle diameter as 1 µm, which is the typical thickness of the platelets. The increase of the yield strength is then estimated to be of 18 MPa. Although this is a rough approximation from the calculation side, it gives a value one order of magnitude below what is experimentally measured, suggesting that the interaction between dislocations and precipitates is not responsible for the yield strength increase. Besides, it is presently shown that α precipitates display noticeable plasticity, and so are not un-shearable particles. This can be a reason for the good ductility of the alloy (no local high stress concentration leading to failure at the precipitate/matrix interface), and may also explain why the hardening is not higher.

The plastic curve of the dual phase alloy shows two very distinct stages, the first one going up to about 10% strain, and the second one from 10% on. In contrast, the reference alloy has a more monotonous tensile evolution. The plateau in the first stage of the dual phase alloy tensile curve is classically attributed to the martensitic transformation [34]. This translates into a very marked hump in the work-hardening rate curve of the dual-phase alloy. It can be noticed that the martensitic plateau is found at 800 MPa, which is rather surprising, since strain-induced martensitic transformation is usually observed at much lower stress below 600 MPa, a value that has long been considered to be one of the limitations in the yield strength. This suggests either that, for the dual phase alloy, the martensitic effect is stronger, or that the twinning is not so significant in the first stage, and does not shield the martensitic plateau on the curve. Recent results showed that martensite may be found as a primary deformation mechanism, taking the shape of needles, or as secondary deformation mechanism where it is rather found at highly strained locations, such as twin/twin or twin/matrix interfaces and intersections, or at grain boundaries [14]. The increased number of interfaces in this alloy compared to the reference one, considering all the α/β matrix interfaces and the α/twin, could explain that the amount of martensite formed as a secondary deformation mechanism is large. Twinning is also possibly not as massive as in the reference alloy in the first stages of the deformation, by comparing microscopy images of Brozek *et al.* [4] and the amount of twins found in this alloy by EBSD, at a similar strain level.

Finally, the work-hardening remains really good, although slightly lower than that of the reference alloy. Classical theories for understanding the TRIP/TWIP alloys work hardening are the dynamic Hall-Petch effect and the composite effect [46,47]. These two effects, the reduction of the dislocation mean free path by formation of interfaces and the necessity of a co-deformation of various domains with different properties, also apply here since both TRIP and TWIP have been confirmed in the alloy. The composite effect should actually be even larger, due to the precipitation of α, possibly inducing an additional mechanical contrast between phases.  The slightly weaker work-hardening may be explained by a decrease capability for twinning in the dual-phase alloy, but, as discussed in the section 4.2, a more thorough understanding of the mechanism by which the twins are interacting with the precipitates is needed to proceed further. Further work consisting in decreasing the coherency of the interfaces between the precipitates and the matrix will also be done, to make it more difficult for the twins to cross the precipitates and increase the hardening. A possibility could be to form spheroidal precipitates for instance, as suggested Lee *et al*., although the interaction twin/martensite/precipitates in this case still has to be investigated [28].

## 5. Conclusion

An approach based on precipitation strengthening is proposed in TRIP/TWIP Ti-alloys to increase the yield strength. An alloy composition is designed based on Calphad thermodynamical predictions and Bo-Md alloy design, to target a dual phase TRIP/TWIP alloy: the β matrix displays TRIP/TWIP properties, to maintain excellent ductility and work-hardening rate, while the precipitates in the matrix bring additional yield strength. The Ti – 7 Cr – 8.5 Sn (wt.%) composition is proposed. The alloy was prepared and the design strategy is successful: the dual phase alloy containing 20% of α precipitates displays both TRIP and TWIP mechanisms, ensuring a uniform elongation of 0.29 and a good work-hardening rate, while its yield strength is 760 MPa, about 200 MPa above that of the reference alloy that has a full β microstructure and the same matrix composition. The precipitation route is thus very effective to increase the critical shear stress to trigger the TRIP and TWIP mechanisms. The impact of the intragranular α precipitates on the deformation mechanisms is more complex. Based on the mechanical behavior, it seems that the TRIP effect is stronger in the first stage for this dual phase alloy than in the reference one, or that the TWIP effect may be lower for the dual phase alloy. However, the presence of the precipitates does not hinder the twinning and martensitic transformation. Detailed investigation of the orientation relationships between the matrix, the twins and the precipitates shows a wide range of behaviors, yet with the main common point that all the precipitates standing in the path of a deformation twin are sheared. The exact mechanisms by which the deformation twins and martensite interacts with the precipitates remain unclear, and more work will be dedicated in understanding that aspect, as well as evaluating the exact contribution of the α precipitates to the mechanical behavior. Overall, these new results open the way to a new branch of phase-to-phase alloy development, where each phase can be tuned individually to bring specific properties to the alloy as a whole.

**Acknowledgements**

The French national research foundation is acknowledged for the funding through the ANR Titwip. M. Garcia is acknowledged for his help with the Image J software.

The alloy design was developed by L.L and F.P., with the help of J-M. J. for the Calphad part. Sample preparation and mechanical testing was done by L.L., Y. D. and L.P. Synchrotron measurements were performed by R. P. under the supervision of R. S. C. Mocuta is acknowledged for his help about synchrotron data treatments. Analysis of the synchrotron data was done by R. P, F. P and R. G. The EBSD analyses was done by L.L, and interpretation of the mechanical and microstructural data was done by L. L., with the help of R.P, Y. D, P.V. and F. P. Manuscript drafting, including the figures, was done by L.L.

**References**

[1] M. Marteleur, F. Sun, T. Gloriant, P. Vermaut, P.J. Jacques, F. Prima, On the design of new β-metastable titanium alloys with improved work hardening rate thanks to simultaneous TRIP and TWIP effects, Scr. Mater. 66 (2012) 749–752. doi:http://dx.doi.org/10.1016/j.scriptamat.2012.01.049.

[2] F. Sun, J.Y. Zhang, M. Marteleur, C. Brozek, E.F. Rauch, M. Veron, P. Vermaut, P.J. Jacques, F. Prima, A new titanium alloy with a combination of high strength, high strain hardening and improved ductility, Scr. Mater. 94 (2015) 17–20. doi:http://dx.doi.org/10.1016/j.scriptamat.2014.09.005.

[3] F. Sun, J.Y. Zhang, C. Brozek, M. Marteleur, M. Veron, E. Rauch, T. Gloriant, P. Vermaut, C. Curfs, P.J. Jacques, F. Prima, The Role of Stress Induced Martensite in Ductile Metastable Beta Ti-alloys Showing Combined TRIP/TWIP Effects, Mater. Today Proc. 2 (2015) S505–S510. doi:https://doi.org/10.1016/j.matpr.2015.07.336.

[4] C. Brozek, F. Sun, P. Vermaut, Y. Millet, A. Lenain, D. Embury, P.J. Jacques, F. Prima, A β-titanium alloy with extra high strain-hardening rate: Design and mechanical properties, Scr. Mater. 114 (2016) 60–64. doi:http://dx.doi.org/10.1016/j.scriptamat.2015.11.020.

[5] D. Banerjee, J.C. Williams, Perspectives on Titanium Science and Technology, Acta Mater. 61 (2013) 844–879. doi:https://doi.org/10.1016/j.actamat.2012.10.043.

[6] W. Wang, X. Zhang, J. Sun, Phase stability and tensile behavior of metastable β Ti-V-Fe and Ti-V-Fe-Al alloys, Mater. Charact. 142 (2018) 398–405. doi:https://doi.org/10.1016/j.matchar.2018.06.008.

[7] A. Paradkar, S.V. Kamat, A.K. Gogia, B.P. Kashyap, Effect of Al and Nb on the trigger stress for stress-induced martensitic transformation during tensile loading in Ti–Al–Nb alloys, Mater. Sci. Eng. A. 487 (2008) 14–19. doi:https://doi.org/10.1016/j.msea.2007.10.021.

[8] A.G. Paradkar, S.V. Kamat, A.K. Gogia, B.P. Kashyap, Various stages in stress–strain curve of Ti–Al–Nb alloys undergoing SIMT, Mater. Sci. Eng. A. 456 (2007) 292–299. doi:https://doi.org/10.1016/j.msea.2006.11.156.

[9] S. Banumathy, R.K. Mandal, A.K. Singh, Structure of orthorhombic martensitic phase in binary Ti–Nb alloys, J. Appl. Phys. 106 (2009). doi:doi:http://dx.doi.org/10.1063/1.3255966.

[10] M. Hida, E. Sukedai, C. Henmi, K. Sakaue, H. Terauchi, Stress induced products and ductility due to lattice instability of β phase single crystal of Ti-Mo alloys, Acta Metall. 30 (1982) 1471–1479. doi:https://doi.org/10.1016/0001-6160(82)90167-5.

[11] L.-F. Huang, B. Grabowski, J. Zhang, M.-J. Lai, C.C. Tasan, S. Sandlöbes, D. Raabe, J. Neugebauer, From electronic structure to phase diagrams: A bottom-up approach to understand the stability of titanium–transition metal alloys, Acta Mater. 113 (2016) 311–319. doi:https://doi.org/10.1016/j.actamat.2016.04.059.

[12] Q. Wang, C. Dong, P.K. Liaw, Structural Stabilities of β-Ti Alloys Studied Using a New Mo Equivalent Derived from [β/(α + β)] Phase-Boundary Slopes, Metall. Mater. Trans. A. 46 (2015) 3440–3447. doi:10.1007/s11661-015-2923-3.

[13] M. Abdel-Hady, K. Hinoshita, M. Morinaga, General approach to phase stability and elastic properties of β-type Ti-alloys using electronic parameters, Scr. Mater. 55 (2006) 477–480. doi:http://dx.doi.org/10.1016/j.scriptamat.2006.04.022.

[14] L. Lilensten, Y. Danard, C. Brozek, S. Mantri, P. Castany, T. Gloriant, P. Vermaut, F. Sun, R. Banerjee, F. Prima, On the heterogeneous nature of deformation in a strain-transformable beta metastable Ti-V-Cr-Al alloy, Acta Mater. 162 (2019) 268–276. doi:https://doi.org/10.1016/j.actamat.2018.10.003.

[15] P. Castany, T. Gloriant, F. Sun, F. Prima, Design of strain-transformable titanium alloys, Comptes Rendus Phys. 19 (2018) 710–720. doi:https://doi.org/10.1016/j.crhy.2018.10.004.

[16] S. Sadeghpour, S.M. Abbasi, M. Morakabati, A. Kisko, L.P. Karjalainen, D.A. Porter, On the compressive deformation behavior of new beta titanium alloys designed by d-electron method, J. Alloys Compd. 746 (2018) 206–217. doi:https://doi.org/10.1016/j.jallcom.2018.02.212.

[17] M.J. Lai, C.C. Tasan, D. Raabe, On the mechanism of 332 twinning in metastable β titanium alloys, Acta Mater. 111 (2016) 173–186. doi:https://doi.org/10.1016/j.actamat.2016.03.040.

[18] N. Hansen, Hall–Petch relation and boundary strengthening, Scr. Mater. 51 (2004) 801–806. doi:https://doi.org/10.1016/j.scriptamat.2004.06.002.

[19] W.L. Wang, X.L. Wang, W. Mei, J. Sun, Role of grain size in tensile behavior in twinning-induced plasticity β Ti-20V-2Nb-2Zr alloy, Mater. Charact. 120 (2016) 263–267. doi:https://doi.org/10.1016/j.matchar.2016.09.016.

[20] S. Sadeghpour, S.M. Abbasi, M. Morakabati, L.P. Karjalainen, D.A. Porter, Effect of cold rolling and subsequent annealing on grain refinement of a beta titanium alloy showing stress-induced martensitic transformation, Mater. Sci. Eng. A. 731 (2018) 465–478. doi:https://doi.org/10.1016/j.msea.2018.06.050.

[21] S. Sadeghpour, S.M. Abbasi, M. Morakabati, A. Kisko, L.P. Karjalainen, D.A. Porter, A new multi-element beta titanium alloy with a high yield strength exhibiting transformation and twinning induced plasticity effects, Scr. Mater. 145 (2018) 104–108. doi:https://doi.org/10.1016/j.scriptamat.2017.10.017.

[22] L. Ren, W. Xiao, C. Ma, R. Zheng, L. Zhou, Development of a high strength and high ductility near β-Ti alloy with twinning induced plasticity effect, Scr. Mater. 156 (2018) 47–50. doi:https://doi.org/10.1016/j.scriptamat.2018.07.012.

[23] J. Zhang, J. Li, G. Chen, L. Liu, Z. Chen, Q. Meng, B. Shen, F. Sun, F. Prima, Fabrication and characterization of a novel β metastable Ti-Mo-Zr alloy with large ductility and improved yield strength, Mater. Charact. (2018). doi:https://doi.org/10.1016/j.matchar.2018.03.031.

[24] L. Lilensten, J.-P. Couzinié, J. Bourgon, L. Perrière, G. Dirras, F. Prima, I. Guillot, Design and tensile properties of a bcc Ti-rich high-entropy alloy with transformation-induced plasticity, Mater. Res. Lett. 5 (2017) 110–116. doi:10.1080/21663831.2016.1221861.

[25] F. Sun, J.Y. Zhang, P. Vermaut, D. Choudhuri, T. Alam, S.A. Mantri, P. Svec, T. Gloriant, P.J. Jacques, R. Banerjee, F. Prima, Strengthening strategy for a ductile metastable β-titanium alloy using low-temperature aging, Mater. Res. Lett. 5 (2017) 547–553. doi:10.1080/21663831.2017.1350211.

[26] J. Gao, A.J. Knowles, D. Guan, W.M. Rainforth, ω phase strengthened 1.2GPa metastable β titanium alloy with high ductility, Scr. Mater. 162 (2019) 77–81. doi:https://doi.org/10.1016/j.scriptamat.2018.10.043.

[27] M. Ahmed, D. Wexler, G. Casillas, O.M. Ivasishin, E.V. Pereloma, The influence of β phase stability on deformation mode and compressive mechanical properties of Ti–10V–3Fe–3Al alloy, Acta Mater. 84 (2015) 124–135. doi:https://doi.org/10.1016/j.actamat.2014.10.043.

[28] S.W. Lee, C.H. Park, J.-K. Hong, J.-T. Yeom, Development of sub-grained α+β Ti alloy with high yield strength showing twinning- and transformation-induced plasticity, J. Alloys Compd. (2019) 152102. doi:https://doi.org/10.1016/j.jallcom.2019.152102.

[29] Q. Xue, Y.J. Ma, J.F. Lei, R. Yang, C. Wang, Mechanical properties and deformation mechanisms of Ti-3Al-5Mo-4.5V alloy with varied β phase stability, J. Mater. Sci. Technol. 34 (2018) 2507–2514. doi:https://doi.org/10.1016/j.jmst.2018.04.004.

[30] Y. Gao, C. Guo, C. Li, Z. Du, Thermodynamic description of the Cr–Sn–Ti system, J. Alloys Compd. 498 (2010) 130–138. doi:https://doi.org/10.1016/j.jallcom.2010.03.140.

[31] S. Basolo, J.F. Bérar, N. Boudet, P. Breugnon, B. Chantepie, J.C. Clémens, P. Delpierre, B. Dinkespiler, S. Hustache, K. Medjoubi, M. Ménouni, C. Morel, P. Pangaud, E. Vigeolas, A 20kpixels CdTe photon-counting imager using XPAD chip, Nucl. Instrum. Methods Phys. Res. Sect. Accel. Spectrometers Detect. Assoc. Equip. 589 (2008) 268–274. doi:https://doi.org/10.1016/j.nima.2008.02.042.

[32] C. Le Bourlot, P. Landois, S. Djaziri, P.-O. Renault, E. Le Bourhis, P. Goudeau, M. Pinault, M. Mayne-L'Hermite, B. Bacroix, D. Faurie, O. Castelnau, P. Launois, S. Rouzière, Synchrotron X-ray diffraction experiments with a prototype hybrid pixel detector, J. Appl. Crystallogr. 45 (2012) 38–47. doi:10.1107/S0021889811049107.

[33] M. Morinaga, Alloy Design Based on Molecular Orbital Method, Mater. Trans. 57 (2016) 213–226. doi:10.2320/matertrans.M2015418.

[34] P. Castany, A. Ramarolahy, F. Prima, P. Laheurte, C. Curfs, T. Gloriant, In situ synchrotron X-ray diffraction study of the martensitic transformation in superelastic Ti-24Nb-0.5N and Ti-24Nb-0.5O alloys, Acta Mater. 88 (2015) 102–111. doi:http://dx.doi.org/10.1016/j.actamat.2015.01.014.

[35] M.J. Lai, C.C. Tasan, J. Zhang, B. Grabowski, L.F. Huang, D. Raabe, Origin of shear induced β to ω transition in Ti–Nb-based alloys, Acta Mater. 92 (2015) 55–63. doi:https://doi.org/10.1016/j.actamat.2015.03.040.

[36] J.M. Silcock, An X-ray examination of the to phase in TiV, TiMo and TiCr alloys, Acta Metall. 6 (1958) 481–493. doi:https://doi.org/10.1016/0001-6160(58)90111-1.

[37] H. Xing, J. Sun, Mechanical twinning and omega transition by ⟨111⟩ 112 shear in a metastable β titanium alloy, Appl. Phys. Lett. 93 (2008) 031908. doi:10.1063/1.2959183.

[38] Y. Yang, P. Castany, E. Bertrand, M. Cornen, J.X. Lin, T. Gloriant, Stress release-induced interfacial twin boundary ω phase formation in a β type Ti-based single crystal displaying stress-induced α" martensitic transformation, Acta Mater. 149 (2018) 97–107. doi:https://doi.org/10.1016/j.actamat.2018.02.036.

[39] S.Q. Wu, D.H. Ping, Y. Yamabe-Mitarai, W.L. Xiao, Y. Yang, Q.M. Hu, G.P. Li, R. Yang, 112〈111〉 Twinning during ω to body-centered cubic transition, Acta Mater. 62 (2014) 122–128. doi:https://doi.org/10.1016/j.actamat.2013.09.040.

[40] N. Stanford, M.R. Barnett, Effect of particles on the formation of deformation twins in a magnesium-based alloy, Mater. Sci. Eng. A. 516 (2009) 226–234. doi:https://doi.org/10.1016/j.msea.2009.04.001.

[41] J.B. Clark, Transmission electron microscopy study of age hardening in a Mg-5 wt.% Zn alloy, Acta Metall. 13 (1965) 1281–1289. doi:https://doi.org/10.1016/0001-6160(65)90039-8.

[42] P. Castany, Y. Yang, E. Bertrand, T. Gloriant, Reversion of a Parent {130}⟨310⟩α'' Martensitic Twinning System at the Origin of {332}<113>β Twins Observed in Metastable β Titanium Alloys, Phys Rev Lett. 117 (2016) 245501. doi:10.1103/PhysRevLett.117.245501.

[43] H. Tobe, H.Y. Kim, T. Inamura, H. Hosoda, S. Miyazaki, Origin of 332 twinning in metastable β-Ti alloys, Acta Mater. 64 (2014) 345–355. doi:https://doi.org/10.1016/j.actamat.2013.10.048.

[44] T. Gladman, Precipitation hardening in metals, Mater. Sci. Technol. 15 (1999) 30–36. doi:10.1179/026708399773002782.

[45] C. Zwikker, Physical properties of solid materials, Pergamon Press, London, 1954.

[46] I. Gutierrez-Urrutia, D. Raabe, Dislocation and twin substructure evolution during strain hardening of an Fe–22wt.% Mn–0.6wt.% C TWIP steel observed by electron channeling contrast imaging, Acta Mater. 59 (2011) 6449–6462. doi:https://doi.org/10.1016/j.actamat.2011.07.009.

[47] O. Bouaziz, N. Guelton, Modelling of TWIP effect on work-hardening, Mater. Sci. Eng. A. 319–321 (2001) 246–249. doi:https://doi.org/10.1016/S0921-5093(00)02019-0.